\documentclass[11pt, onecolumn]{IEEEtran}
\usepackage{amsmath,amssymb,euscript,yfonts,psfrag,latexsym,dsfont,graphicx}
\usepackage{bbm,color,amstext,wasysym,subfig,flushend,parskip,balance}
\graphicspath{{./},{./figures/}}

\newtheorem{thm}{Theorem}

\newtheorem{prop}{Proposition}
\newtheorem{problem}[thm]{Problem}
\newtheorem{remark}[thm]{Remark}

\newcommand{\mR}{{\mathbb R}}

\newcommand{\mE}{{\mathbb E}}
\newcommand{\cN}{{\mathcal N}}

\newcommand{\f}{{\mathfrak f}}

\newcommand{\xmu}{{\bar x_\mu}}
\newcommand{\range}{\operatorname{range}}

\newcommand{\trace}{\operatorname{tr}}
\newcommand{\argmin}{\operatorname{argmin}}

\definecolor{grey}{rgb}{0.6,0.6,0.6}
\definecolor{lightgray}{rgb}{0.97,.99,0.99}

\begin{document}
\title{Steering the distribution of agents \\in mean-field and cooperative games}

\author{Yongxin Chen, Tryphon Georgiou and Michele Pavon
\thanks{Y.~Chen is with the
Department of Electrical and Computer Engineering, Iowa State University, IA 50011,
Ames, Iowa, USA; {yongchen@iastate.edu}}
\thanks{T.T.\ Georgiou is with the Department of Mechanical and Aerospace Engineering,
University of California, Irvine, CA 92697; {tryphon@uci.edu}}
\thanks{M.\ Pavon is with the Dipartimento di Matematica ``Tullio Levi Civita",
Universit\`a di Padova, via Trieste 63, 35121 Padova, Italy; {pavon@math.unipd.it}}
\thanks{
Supported in part by the
NSF under grant ECCS-1509387,
the AFOSR under grants FA9550-15-1-0045 and FA9550-17-1-0435, and by the University of Padova Research Project CPDA 140897.}}
\markboth{\today}{}

\maketitle

\begin{abstract}
The purpose of this work is to pose and solve the problem to guide a collection of weakly interacting dynamical systems (agents, particles, etc.) to a specified terminal distribution. The framework is that of mean-field and of cooperative games. A terminal cost is used to accomplish the task; we establish that the map between terminal costs and terminal probability distributions is onto. Our approach relies on and extends the theory of optimal mass transport and its generalizations.
\end{abstract}


\section{Introduction}
Mean-field game (MFG) theory is the study of noncooperative games involving a large number of agents.
The basic model requires agents to abide by identical dynamics and seek to minimize an individual cost function that is also the same for all. As the number of agents increases to infinity, the empirical distribution of their states becomes indifferent to the strategy of any single agent and yet it couples their individual responses. Thus, the aggregate response of the agents (mean field) drives individual responses while the action of individual agents is insignificant.
On the flip side, cooperative games refer to the situation where agents seek to jointly optimize a common performance index. Either way, the desire to minimize cost, individually or collectively, drives the empirical distribution of agents in suitable ways.
The purpose of this work to study for both, MFG and cooperative games, the control problem to steer the collective response of agents over a finite window of time between two specified end-point marginal distributions by suitable choice of cost (i.e., incentives) in non-cooperative games and centralized control with cooperative agents, and also the problem to ensure a desired stationary distribution under similar conditions. This viewpoint is influenced by optimal mass transport (OMT) theory that deals with the flow of time-marginal densities for a collective (agents, particles, resources) and corresponding control and modeling problems.

The study of MFG's was introduced into the engineering literature by Huang, Malham\'e and Caines \cite{HuaMalCai06} and, independently, by Lasry and Lions \cite{LasLio07}. Earlier, in the economics literature, similar models were considered by Jovanovic and Rosental \cite{JovRos88}. The importance of the subject stems from the wide range of applications that include modeling and control of multi-agent dynamical systems, stock market dynamics, crowd dynamics, power systems and more; see \cite{HuaMalCai06,LasLio07,NouCai13,CarDelLac13}, and also see \cite{HuaCaiMal03,HuaCaiMal07,BenSunYamYun14,Bar12} in the special case of linear dynamics and quadratic cost.
On the other hand, OMT originates in the work of Monge \cite{Mon81} and aims directly at relating/transporting distributions under minimum cost.  Kantorovich \cite{Kan42} introduced linear programming and duality theory for solving OMT resource allocation problems and, in recent years, a fast developing phase was spurred by a wide range of applications of OMT to probability theory, economics, weather modeling, biology and mathematical physics 
 \cite{GanMcc96,evans1997partial,EvaGan99,
Vil03,AmbGigSav06,Vil08,ambrosio2013user,santambrogio2015optimal}. The connection between dynamic OMT \cite{BenBre00} and stochastic control has been explored in our work, e.g.\ \cite{CheGeoPav14e,CheGeoPav15b}, where the focus has been on regulating stochastic uncertainty of diffusion processes and of stochastically driven dynamical systems by suitable control action. These stochastic control problems, in turn, relate to a classical maximum entropy problem on path space known as the Schr\"odinger bridge problem, see e.g., \cite{Dai91,Leo12,Leo13,GenLeoRip15,CheGeoPav14a,CheGeoPav14b}.
 
The goal of the present work is to study density steering problems in an MFG framework or, equivalently, explore the role of interaction potential and decentralized strategies when steering an initial distribution to a terminal one. In particular, we are interested on how to design an added terminal cost so as to provide incentives for agents, under a Nash equilibrium strategy, to move collectively as specified. To this end, we establish that the map between terminal costs and terminal probability distributions is onto. Thereby, we develop an MFG-based synthesis framework for OMT-type stochastic control problems with or without stochastic excitation.

The paper evolves along the following lines. First, we discuss the motivation and problem formulation in Section \ref{sec:formulation}. The solution is provided in Section \ref{sec:approach}. In Section  \ref{sec:zeronoise} we study similar problems with less or no disturbance. Section \ref{sec:gaussian} is dedicated to the special case with Gaussian marginal distributions. Similar problems in the stationary setting are investigated in Section \ref{sec:stationary}. In Section \ref{sec:coop}, we developed the cooperative game counterpart of the density steering problem. This follows by a simple academic example in Section \ref{sec:example} and a brief concluding remark in Section \ref{sec:conclusion}.

\newcommand{\barA}{{\bar{A}}}
\section{Problem formulation}\label{sec:formulation}
We herein investigate the collective dynamical response of a group of agents (also thought of as particles, players, and so on) that interact weakly with each other. The terminology ``weakly'' refers to the agents being statistically indistinguishable (anonymous) and affecting each other's response only through their empirical distribution \cite{Fis14}.
Thus, we consider such a system of $N$ agents with dynamics\footnote{This type of weakly coupled system of linear stochastic models has been studied in  \cite{HuaCaiMal03,HuaCaiMal07,BenSunYamYun14}. In our setting we further assume that the noise $dw_i$ and control action $u$ affect the dynamics in a similar manner, through the same matrix $B$. The more general case, where this is not so, is more demanding and will be pursued in future publication, cf.\ \cite{CheGeoPav14a,CheGeoPav14b}.}
specified by
	\begin{eqnarray}\label{eq:dynagent}
		dx_i(t)&=&Ax_i(t)dt+\frac{1}{N-1}\sum_{j\neq i} \barA x_j(t)dt+Bu_i(t)dt+Bdw_i(t),\\\nonumber &&~~x_i(0)=x_0^i,~~i=1,\ldots,N.
	\end{eqnarray}
Here, $x_i, u_i, w_i$ represent the state, control input, white noise disturbance, respectively, for the $i$th agent, and the model parameters are the same for all. We further assume that their initial conditions $x_0^1, x_0^2,\ldots,x_0^N$ are all independent with the same probability density $\rho_0$. The $i$th agent interacts with the rest through the averaged position.
The matrices $A,\bar A\in \mR^{n\times n}, B\in \mR^{n\times m}$ are continuous functions of time; for notational simplicity we often use e.g., $A$ instead of $A(t)$. It is assumed that the pair is controllable in the sense that the reachability Gramian 
	\[
		M(t,s)=\int_s^t \Phi(t,\tau) B(\tau)B(\tau)'\Phi(t,\tau)'d\tau
	\]
is invertible for all $s<t$. Here, $\Phi(\cdot,\cdot)$ denotes the state transition matrix that is defined via
	\[
		\frac{\partial\Phi(t,s)}{\partial t}=A\Phi(t,s), ~~\Phi(s,s)=I.
			\]
Clearly, in case when $A$ is time-invariant, $\Phi(t,s)=e^{A(t-s)}$.

In MFG \cite{HuaMalCai06}, each agent searches for an optimal control strategy to minimize its own cost\footnote{For simplicity of notation and without loss in generality we take the end point to be $t=1$.}
	\begin{equation}\label{eq:cost}
		J_i(u_i)=\mE\left\{\int_0^1 f(t,x_i(t),u_i(t),\mu^N(t)) dt+g(x_i(1),\mu^N(1))\right\},
	\end{equation}
where
	\begin{equation}\label{eq:empiricaldis}
		\mu^N(t) =\frac{1}{N}\sum_{i=1}^N \delta_{x_i(t)}
	\end{equation}
is the empirical distribution of the states of the $N$ agents at time $t$. Thus, this is a non-cooperative game and the cost of the $i$th agent is affected by the strategies of others only through the empirical distribution. An optimal control corresponds to a Nash equilibrium for the game.
We follow the arguments in \cite{CarDelLac13}, and restrict ourselves to equilibria that correspond to symmetric Markovian control strategies (state feedback)
	\begin{equation}\label{eq:symstrategy}
		u_i(t)=\phi(t,x_i(t)), i=1,\ldots,N.
	\end{equation}
	
When $N$ is large, the empirical distribution $\mu^N$ is indifferent to small perturbations of control strategy of a single agent. This points to the following approach  \cite{CarDelLac13} to obtain an approximate Nash equilibrium: fix a family $(\mu(t))_{0\le t\le 1}$ of probability measures and solve the standard stochastic control problem  
	\begin{equation}\label{eq:optimalcontrol}
		\phi^\star=\argmin_{\phi} \mE\left\{\int_0^1 f(t,x(t),\phi(t,x(t)),\mu(t)) dt+g(x(1),\mu(1))\right\}
	\end{equation}
subject to the dynamics
	\begin{equation}\label{eq:dynamics}
		dx(t)=Ax(t)dt+\barA\xmu(t)dt +B\phi(t,x(t))dt+Bdw(t), ~~x(0)=x_0~ \mbox{a.s.}
	\end{equation}
where
\[
\bar x_\mu := 
\langle x, \mu(t)\rangle
\]
denotes the mean\footnote{Throughout, we use the expressions $\bar x_\mu$ or $\langle x, \mu(t)\rangle$ interchangeably.} of the distribution $\mu(t)$, and $x_0$ is a random vector with probability density $\rho_0$. 
Considering the choice $(\mu(t))_{0\le t\le 1}$ as a parameter, the remaining issue is to choose this distribution flow so that the actual distribution of the solution $x(t)$ of \eqref{eq:dynamics} with optimal control strategy 
	\begin{equation}\label{eq:nashcontrol}
		u^\star(t)=\phi^\star(t,x(t))
	\end{equation}
coincides with $\mu(t)$.
The solution to the MFG problem involves establishing the existence and uniqueness of the solution to two coupled partial differential equations (PDEs) \cite{CarDelLac13}. 
It has been shown that a Nash equilibrium point for this mean-field game exists under rather mild assumptions on the cost function \cite{HuaMalCai06,LasLio07,NouCai13,CarDelLac13,HuaCaiMal03,HuaCaiMal07,BenSunYamYun14,Bar12}. That is, there exists a family $(\mu(t))_{0\le t\le 1}$ such that the distribution flow of the solution $x(t)$ of \eqref{eq:dynamics} under optimal control strategy $\phi^\star$ coincides with this same $\mu$. In addition, this optimal control $\phi^\star$ is proven to be an $\varepsilon$-Nash equilibrium to the $N$-player-game for $N$ large \cite{CarDelLac13,Car10}. 
	  
Departing from previous literature, this paper deals with the density steering problem of the $N$-player-game system. More specifically, we are interested in introducting a suitable cost incentive so that the system is driven to a specific distribution $\mu_1$ at time $t=1$ under \eqref{eq:nashcontrol}. In fact, it turns out that under mild conditions, a quadratic running cost in both the control and state (i.e., group linear tracking as in the work of Huang, Malam\'e and Caines \cite{HuaMalCai06}), can be enhanced by a suitable terminal cost $g$ as follows
	\begin{equation}\label{eq:cost1}
	J_i(u_i)=\mE\left\{\int_0^1 \left(\frac{1}{2}\|u_i(t)\|^2 +\frac12\|x_i(t)-\bar x(t)\|_Q^2\right) dt+g(x_i(1),\mu^N(1))\right\}
	\end{equation}
so as to acomplish the task of steering the initial distribution to the desired terminal one. In other words, we show that the mapping between a choice of $g$ and the terminal distribution $\mu_1$ is onto.
Formally, the problem we are interested in can be stated as follows.
\begin{problem}\label{pro:meanfield}
Given $N$ agents governed by \eqref{eq:dynagent} with initial probability density $\rho_0$, find a terminal cost $g$ such that, in the Nash equilibrium with cost functional \eqref{eq:cost1}, the agents will reach a given terminal density $\rho_1$ at time $t=1$, in the limit as $N$ goes to $\infty$.
\end{problem}

\section{General approach and solution}\label{sec:approach}
Without loss of generality and for simplicity of the exposition we only consider a running cost only in the control actuation (i.e., taking the matrix $Q$ in \eqref{eq:cost1} to be zero).
We begin by considering the optimal steering problem \cite{Che16,CheGeoPav14a,CheGeoPav14b,CheGeoPav17c}
without terminal cost, i.e., 
for a fixed density flow $(\mu(t))_{0\le t\le 1}$, consider the control problem to minimize 
	\[
		J(u)=\mE\left\{\int_0^1 \frac{1}{2}\|u(t)\|^2 dt\right\}
	\]
subject to the dynamics
	\begin{equation}\label{eq:dynamicsnocontrol}
		dx(t)=Ax(t)dt+\barA\xmu(t)dt +Bu(t)dt+Bdw(t), ~~x(0)=x_0 \sim \rho_0~ 
	\end{equation}
and the constraint that $x(1)$ has probability density $\rho_1$. This problem can be (formally) posed as
	\begin{subequations}\label{eq:optimization}
    \begin{align}
    & \inf_{\rho, u}\quad \int_0^1 \int_{\mR^n} \frac{1}{2} \rho(t,x) \|u(t,x)\|^2 dxdt, \\
    & \frac{\partial \rho}{\partial t}+\nabla\cdot((Ax+\barA\xmu+Bu)\rho)-\frac{1}{2}\trace(BB'\nabla^2 \rho)=0, \\
    & \rho(0,\cdot)=\rho_0, \quad \rho(1,\cdot)=\rho_1\label{eq:optimization3}.
    \end{align}
    \end{subequations}
Following a similar argument as in \cite{CheGeoPav15b}, we can establish the following sufficient condition for optimality.
\begin{prop}\label{prop:necessary}
If there exists a function $\lambda$ such that $\rho^\star, \lambda$ satisfy
    \begin{subequations}\label{eq:necessary}
    \begin{equation}\label{eq:necessary1}
        \frac{\partial \lambda}{\partial t}+\nabla\lambda\cdot Ax+\nabla\lambda\cdot\barA\xmu+\frac{1}{2}\trace(BB'\nabla^2 \lambda)+\frac{1}{2}\nabla\lambda\cdot BB'\nabla\lambda=0,
    \end{equation}
    \begin{equation}\label{eq:necessary2}
    \frac{\partial \rho^\star}{\partial t}+\nabla\cdot((Ax+\barA\xmu+BB'\nabla\lambda)\rho^\star)-\frac{1}{2}\trace(BB'\nabla^2 \rho^\star)=0,
    \end{equation}
    and boundary conditions
    \begin{equation}\label{eq:necessary3}
    \rho^\star(0,\cdot)=\rho_0, \quad \rho^\star(1,\cdot)=\rho_1,
    \end{equation}
    \end{subequations}
then $(\rho^\star, u^\star=B'\nabla\lambda)$ is a solution to \eqref{eq:optimization}.
\end{prop}

Replacing $\mu$ in \eqref{eq:necessary} by $\rho^\star$ we obtain the system of (nonlinear) PDE's	   \begin{subequations}\label{eq:necessarynon}
    \begin{equation}\label{eq:necessarynon1}
        \frac{\partial \lambda}{\partial t}+\nabla\lambda\cdot Ax+\nabla\lambda\cdot\barA\bar x_{\rho^\star}+\frac{1}{2}\trace(BB'\nabla^2 \lambda)+\frac{1}{2}\nabla\lambda\cdot BB'\nabla\lambda=0,
    \end{equation}
    \begin{equation}\label{eq:necessarynon2}
    \frac{\partial \rho^\star}{\partial t}+\nabla\cdot((Ax+\barA\bar x_{\rho^\star}+BB'\nabla\lambda)\rho^\star)-\frac{1}{2}\trace(BB'\nabla^2 \rho^\star)=0,
    \end{equation}
    \begin{equation}\label{eq:necessarynon3}
    \rho^\star(0,\cdot)=\rho_0, \quad \rho^\star(1,\cdot)=\rho_1.
    \end{equation}
    \end{subequations}
We will show below that, under mild assumptions, \eqref{eq:necessarynon} has a solution. In fact, we will construct such a solution relying on standard Schr\"odinger bridge theory \cite{CheGeoPav14e,CheGeoPav15b}.

\begin{remark}
Note that the coupled PDEs (\ref{eq:necessarynon1}-\ref{eq:necessarynon2}) are the same as the PDEs that arise in classic MFG problems corresponding to \eqref{eq:dynagent} and \eqref{eq:cost1}. 
However, the usual boundary conditions
    \begin{equation}\label{eq:meanfieldgame3}\nonumber
    \rho^\star(0,\cdot)=\rho_0, \quad \lambda(1,x)=-g(x,\rho^\star(1,\cdot)),
    \end{equation}
are now different and given by \eqref{eq:necessarynon3}. Evidently, the Lagrange multiplier $-\lambda$ is  the value (cost-to-go) function of the associated optimal control problem.
\end{remark}

To solve \eqref{eq:necessarynon}, we first consider the Schr\"odinger bridge problem with prior dynamics
    \begin{equation}\label{eq:dynamics1}
        dx(t)=Ax(t)dt+Bdw(t).
    \end{equation}
Let
    \[
        \hat{\rho}_0(x)=\rho_0(x+\bar x_{\rho_0})
    \]
and
    \[
        \hat{\rho}_1(x)=\rho_1(x+\bar x_{\rho_1}),
    \]
then
    \[
        \langle x, \hat{\rho}_0 \rangle =0, ~~\langle x, \hat{\rho}_1 \rangle=0.
    \]
The Schr\"odinger bridge with prior dynamics \eqref{eq:dynamics1} and marginal distributions $\hat\rho_0$ and $\hat\rho_1$ is \cite{CheGeoPav14e,CheGeoPav15b}
    \begin{equation}\label{eq:SB1}
        dx(t)=Ax(t)dt+BB'\nabla\hat\lambda(t,x(t))dt+Bdw(t),
    \end{equation}
where $\hat\lambda$ satisfies
    \begin{equation}\label{eq:lambdahat}
        \frac{\partial \hat\lambda}{\partial t}+\nabla\hat\lambda\cdot Ax+\frac{1}{2}\trace(BB'\nabla^2\hat\lambda)+\frac{1}{2}\nabla\hat\lambda\cdot BB'\nabla\hat\lambda=0,
    \end{equation}
or, equivalently, $\varphi=\exp(\hat\lambda)$ satisfies
    \[
        \frac{\partial \varphi}{\partial t}+\nabla\varphi\cdot Ax+\frac{1}{2}\trace(BB'\nabla^2\varphi)=0.
    \]
The boundary condition $\hat\lambda(1,\cdot)$ for $\hat\lambda$ is chosen in a way so that the resulting density flow $\hat\rho(t,x)$ of \eqref{eq:SB1}, which is
	\[
		\frac{\partial\hat\rho}{\partial t}+\nabla\cdot((Ax+BB'\nabla\hat\lambda)\hat\rho)
		-\frac{1}{2}\trace(BB'\nabla^2 \hat\rho)=0,
	\]
matches the marginal distributions $\hat\rho_0$ and $\hat\rho_1$. The pair $(\hat\lambda, \hat\rho)$ satisfies that
	\begin{equation}\label{eq:zeromeancontrol}
        \langle \nabla\hat\lambda(t,\cdot), \hat\rho(t,\cdot)\rangle =0
    \end{equation}
and therefore
        \begin{equation}\label{eq:zeromeanflow}
        \langle x, \hat\rho(t,\cdot)\rangle =0
    \end{equation}
for all $0\le t\le 1$. The intuition is that if the expectation of the control, i.e., $\langle \nabla\hat\lambda(t,\cdot), \hat\rho(t,\cdot)\rangle$, is not constantly $0$, then one can always shift the control by its mean to achieve a smaller cost.
Now let
    \[
        m(t)=\Phi(1,t)'\bar M_{10}^{-1}(\bar x_{\rho_1}-\bar\Phi_{10}\bar x{\rho_0}),
    \]
$y(t)$ the solution to
    \[
        \dot{y}(t)=(A+\barA)y(t)+BB'm(t), ~~y(0)=\bar x_{\rho_0},
    \]
and 
	\[
		\gamma(t)=-\int_0^t (\barA y(s)\cdot m(s)+\frac{1}{2}m(s)\cdot BB'm(s))ds.
	\]
Here $\bar\Phi_{10}:=\bar\Phi(1,0)$ with $\bar\Phi$ being the state transition matrices for the pair $(A+\barA , B)$ and the ``coupled'' Gramian
    \[
        \bar M_{10}=\int_0^1 \bar\Phi(1,\tau)BB'\Phi(1,\tau)'d\tau
    \]
is assumed to be invertible. Note that $y(1)=\bar x_{\rho_1}$.

With these ingredients, we construct a solution to \eqref{eq:necessarynon} as follows. Define $(\lambda, \rho^\star)$ by
    \begin{subequations}\label{eq:optimalpair}
    \begin{equation}\label{eq:lambda}
        \lambda(t,x)=\hat\lambda(t,x-y(t))+m(t)\cdot x+\gamma(t),
    \end{equation}
    and
    \begin{equation}\label{eq:rhostar}
        \rho^\star(t,x)=\hat\rho(t,x-y(t)).
    \end{equation}
    \end{subequations}
In so doing, $(\lambda, \rho^\star)$ is a solution of \eqref{eq:necessarynon}. On one hand, substituting \eqref{eq:optimalpair} into \eqref{eq:necessarynon2}, in view of \eqref{eq:zeromeanflow}, we obtain
    \begin{eqnarray*}
        &&\frac{\partial \rho^\star}{\partial t}+\nabla\cdot((Ax+\barA \bar x_{\rho^\star}+BB'\nabla\lambda)\rho^\star)-\frac{1}{2}\trace(BB'\nabla^2 \rho^\star)
        \\&=& \frac{\partial \hat\rho}{\partial t}-\nabla\hat\rho\cdot((A+\barA )y+BB'm)\\&&~~~ +\nabla\cdot((Ax+BB'\nabla\lambda)\hat\rho)+\barA \langle \xi, \hat\rho(t,\xi-y)\rangle\cdot \nabla \hat\rho-\frac{1}{2}\trace(BB'\nabla^2 \hat\rho)
        \\&=&-\nabla\hat\rho\cdot(\barA y+BB'm)+\nabla\cdot (BB'm\hat\rho)+\barA \langle\xi, \hat\rho(t,\xi-y)\rangle\cdot\nabla\hat\rho=0,
    \end{eqnarray*}
where we referred to \eqref{eq:zeromeanflow} in the last step. The fact that $\rho^\star$ matches the boundary conditions \eqref{eq:necessarynon3} follows directly from the definition \eqref{eq:rhostar}. On the other hand, plugging \eqref{eq:optimalpair} into \eqref{eq:necessarynon1} yields
    \begin{eqnarray*}
    &&\frac{\partial \lambda}{\partial t}+\nabla\lambda\cdot Ax+\nabla\lambda\cdot\barA \bar x_{\rho^\star}+\frac{1}{2}\trace(BB'\nabla^2 \lambda)+\frac{1}{2}\nabla\lambda\cdot BB'\nabla\lambda
    \\&=& \frac{\partial \hat\lambda}{\partial t}-\nabla\hat\lambda\cdot((A+\barA )y+BB'm)+\dot{m}\cdot x+\dot{\gamma}\\&&~~~ +(\nabla\hat\lambda+m)\cdot (Ax+\barA \bar x_{\rho^\star})+\frac{1}{2}\trace(BB'\nabla^2 \hat\lambda)
    \\&&+\frac{1}{2}(\nabla\hat\lambda+m)\cdot BB'(\nabla\hat\lambda+m)
    \\&=& -\nabla\hat\lambda\cdot\barA y+\dot{m}\cdot x+\dot{\gamma}+(\nabla\hat\lambda+m)\cdot\barA \bar x_{\rho^\star}+m\cdot Ax+\frac{1}{2}m\cdot BB'm=0.
    \end{eqnarray*}
Therefore $(\lambda, \rho^\star)$ in \eqref{eq:optimalpair} is indeed a solution to \eqref{eq:necessarynon}.
Finally, back to Problem \ref{pro:meanfield}, we assert that with terminal cost
	\begin{equation}\label{eq:terminalcost}
	g(x,\mu)=-\hat\lambda(1,x-\xmu)-m(1)\cdot x-\gamma(1),
	\end{equation}
we can lead the agents to have terminal distribution $\rho_1$. To this extent, we follow the strategy in \cite{CarDelLac13} as mentioned in Section \ref{sec:formulation}. First fix $\mu=\rho^\star$ with $\rho^\star$ as in \eqref{eq:rhostar}, and then solve the optimal control problem \eqref{eq:optimalcontrol}. Since 
	$g(x,\rho^\star(1,\cdot))=g(x,\rho_1)=-\lambda(1,x)$,
we have
	\begin{eqnarray*}
		&&\mE\left\{\int_0^1 \frac{1}{2}\|u(t)\|^2 dt+g(x,\rho^\star(1,\cdot))\right\}
		\\&=&
		\mE\left\{\int_0^1 \frac{1}{2}\|u(t)\|^2 dt-\lambda(1,x(1))\right\}
		\\&=&
		\mE\left\{\int_0^1 [\frac{1}{2}\|u(t)\|^2 dt-d\lambda(t,x(t))]-\lambda(0,x(0))\right\}
		\\&=&
		\mE\left\{\int_0^1 [\frac{1}{2}\|u(t)\|^2 dt-\frac{\partial \lambda}{\partial t}dt-\nabla\lambda\cdot dx(t)-\frac{1}{2}\trace(BB'\nabla^2 \lambda)dt]-\lambda(0,x(0))\right\}
		\\&=&
		\mE\left\{\int_0^1 \frac{1}{2}\|u(t)-B'\nabla\lambda(t,x(t))\|^2 dt\right\}-\mE\{\lambda(0,x(0)\}.
	\end{eqnarray*}
Hence, the unique optimal control strategy is $u^\star(t)=B'\nabla\lambda(t,x(t))$. It follows from \eqref{eq:necessarynon} that the probability distribution of the controlled state $x(t)$ is $\rho^\star$. Therefore, with terminal cost $g$ as in \eqref{eq:terminalcost} we are able to steer the system to terminal distribution $\rho_1$. Thus, we have established the following result.

\begin{thm}
Consider $N$ agents governed by \eqref{eq:dynagent} with initial density $\rho_0$. Suppose the terminal cost in \eqref{eq:cost1} is as in \eqref{eq:terminalcost}. Then, in the Nash equilibrium, the agents will reach density $\rho_1$ at time $t=1$, in the limit as $N$ goes to $\infty$.
\end{thm}
\begin{remark}
In fact, the dependence of $g$ on $\mu$ is not necessary. One can simply take $g(x,\mu)=g(x)=-\lambda(1,x)$. With this terminal cost, we can still conclude that $(\lambda, \rho^\star)$ corresponds to a Nash equilibrium as well. This is due to the fact that we fix the density flow first when we derive a Nash equilibrium. We might need the dependence of $g$ on $\mu$ to conclude the uniqueness of the equilibrium. It is unclear to us if this is the case.
\end{remark}

\section{Zero-noise limit}\label{sec:zeronoise}
In this section, we study the same problem (Problem \ref{pro:meanfield}), with however reduced disturbance. More specifically, we consider a system of $N$ agents with dynamics
	\begin{eqnarray}\label{eq:dynamicsepsilon}
		dx_i(t)&=&Ax_i(t)dt+\frac{1}{N-1}\sum_{j\neq i} \barA x_j(t)dt+Bu_i(t)dt+\sqrt{\epsilon}Bdw_i(t),\\&& ~~x_i(0)=x_0^i,~~i=1,\ldots,N,\nonumber
	\end{eqnarray}
where $\epsilon>0$ represents the variance of the noise. We are especially interested in the limit behavior of the solution to Problem \ref{pro:meanfield} with dynamics \eqref{eq:dynamicsepsilon} when $\epsilon$ goes to $0$. Following the same arguments as in Section \ref{sec:approach}, we arrive at the coupled PDEs 
	   \begin{subequations}\label{eq:sNecessary}
    \begin{equation}\label{eq:snecessary1}
        \frac{\partial \lambda}{\partial t}+\nabla\lambda\cdot Ax+\nabla\lambda\cdot\barA \bar x_{\rho^\star}+\frac{\epsilon}{2}\trace(BB'\nabla^2 \lambda)+\frac{1}{2}\nabla\lambda\cdot BB'\nabla\lambda=0,
    \end{equation}
    \begin{equation}\label{eq:snecessary2}
    \frac{\partial \rho^\star}{\partial t}+\nabla\cdot((Ax+\barA \bar x_{\rho^\star}+BB'\nabla\lambda)\rho^\star)-\frac{\epsilon}{2}\trace(BB'\nabla^2 \rho^\star)=0,
    \end{equation}
    \begin{equation}\label{eq:snecessary3}
    \rho^\star(0,\cdot)=\rho_0, \quad \rho^\star(1,\cdot)=\rho_1.
    \end{equation}
    \end{subequations}
The optimal control strategy is given by $u(t)=B'\nabla\lambda(t,x(t))$ and terminal cost $g$ is as in \eqref{eq:terminalcost} with adjusted diffusitivity. 

Taking the limit of \eqref{eq:sNecessary} as $\epsilon\rightarrow 0$ gives
\begin{subequations}\label{eq:Inecessary}
    \begin{equation}\label{eq:lnecessary1}
        \frac{\partial \lambda}{\partial t}+\nabla\lambda\cdot Ax+\nabla\lambda\cdot\barA \bar x_{\rho^\star}+\frac{1}{2}\nabla\lambda\cdot BB'\nabla\lambda=0,
    \end{equation}
    \begin{equation}\label{eq:lnecessary2}
    \frac{\partial \rho^\star}{\partial t}+\nabla\cdot((Ax+\barA \bar x_{\rho^\star}+BB'\nabla\lambda)\rho^\star)=0,
    \end{equation}
    \begin{equation}\label{eq:lnecessary3}
    \rho^\star(0,\cdot)=\rho_0, \quad \rho^\star(1,\cdot)=\rho_1.
    \end{equation}
    \end{subequations}
With similar analysis as in Section \ref{sec:approach} we conclude that the above PDEs system has a (viscosity) solution \cite{FleSon06}. In particular, the solution $(\lambda,\rho^\star)$ to \eqref{eq:Inecessary} has the form \eqref{eq:optimalpair} with $\hat\lambda$ being
	\[
        \hat\lambda(t,x)=\inf_y \left\{\hat\lambda(0,y)+
        \frac{1}{2}(x-\Phi(t,0)y)'M(t,0)^{-1}(x-\Phi(t,0)y)\right\},
 	\]
where
    \begin{equation*}
        \hat\lambda(0,x)=\psi(M_{10}^{-1/2}\Phi_{10}x)-\frac{1}{2}x'\Phi_{10}'M_{10}^{-1}\Phi_{10}x
    \end{equation*}
 and $\psi$ corresponds to the optimal transport map with prior dynamics $\dot{x}=Ax+Bu$, and marginal distributions $\hat{\rho}_0$ and $\hat{\rho}_1$ after coordinate transformation, see \cite[Proposition 2]{CheGeoPav15b}.
 The solution to \eqref{eq:Inecessary} in fact solves the following problem. 
\begin{problem}
Given $N$ agents governed by \eqref{eq:dynamicsepsilon} with $\epsilon=0$, and initial probability density $\rho_0$, find a function $g$ such that, in the Nash equilibrium with cost function \eqref{eq:cost1}, the agents would reach a specified density $\rho_1$ at time $t=1$, in the limit as $N$ goes to $\infty$.
\end{problem}
With the solution to \eqref{eq:Inecessary}, we can choose a terminal cost as in \eqref{eq:terminalcost}. The corresponding equilibrium control strategy is $u(t,x)=B'\nabla\lambda(t,x)$.
 \begin{thm}
Consider $N$ agents governed by \eqref{eq:dynamicsepsilon} with $\epsilon=0$ and initial density $\rho_0$. Suppose the terminal cost $g$ in \eqref{eq:cost1} is as in \eqref{eq:terminalcost}, then, in the Nash equilibrium, the agents will reach density $\rho_1$ at time $t=1$, in the limit as $N$ goes to $\infty$.
 \end{thm}

\section{Gaussian case}\label{sec:gaussian}
In the special case when $\rho_0$ and $\rho_1$ are normal (Gaussian) distributions, the solutions have a nice linear structure. Let the two marginal distributions be 
	\[
		\rho_0 \sim \cN[m_0, \Sigma_0],~~~ \rho_1 \sim \cN[m_1,\Sigma_1],
	\]
i.e., Gaussian distributions with, respectively, means $m_0, m_1$ and covariances $\Sigma_0, \Sigma_1$. When $\epsilon=1$, $\hat\lambda$ in \eqref{eq:lambdahat} equals
	\[
		\hat\lambda(t,x)=-\frac{1}{2}x'\Pi(t)x+\frac{1}{2}\int_0^t \trace(BB'\Pi(s))ds,
	\]
where $\Pi(t)$ is the solution to the Riccati equation
    \begin{equation}\label{eq:ric}
        \dot{\Pi}(t)=-A'\Pi(t)-\Pi(t)A+\Pi(t)BB'\Pi(t)
    \end{equation}
with boundary condition
    \begin{align*}
        \Pi(0)&=\Sigma_0^{-1/2}\left[\frac{I}{2}+\Sigma_0^{1/2}\Phi_{10}'M_{10}^{-1}\Phi_{10}\Sigma_0^{1/2}\right.\\&~~~\left.-(\frac{I}{4}+\Sigma_0^{1/2}\Phi_{10}'M_{10}^{-1}\Sigma_1M_{10}^{-1}\Phi_{10}\Sigma_0^{1/2})^{1/2}\right]\Sigma_0^{-1/2}.
    \end{align*}
where $\Phi_{10}=\Phi(1,0), M_{10}=M(1,0)$. 
And so, in view of \eqref{eq:terminalcost}, one choice of terminal cost is
	\begin{equation}\label{eq:finalcostlinear}
		g(x,\mu)=\frac{1}{2}(x-\xmu)'\Pi(1)(x-\xmu)-m(1)\cdot x.
	\end{equation}
In the above we have discarded some constant terms as it doesn't affect the final result.
\begin{thm}
Consider $N$ agents governed by \eqref{eq:dynagent} with initial density $\rho_0\sim \cN[m_0, \Sigma_0]$. Suppose the terminal cost in \eqref{eq:cost1} is \eqref{eq:finalcostlinear}. Then, in the Nash equilibrium, the agents will reach density $\rho_1\sim \cN[m_1,\Sigma_1]$ at time $t=1$, in the limit as $N$ goes to $\infty$.
\end{thm}

Following the discussion in Section \ref{sec:zeronoise}, the solution to the problem with noise intensity $\epsilon$ is almost identical to the above except that, the initial condition of the Riccati equation \eqref{eq:ric} becomes
	\begin{align*}
		\Pi_\epsilon(0)&=\Sigma_0^{-1/2}\left[\frac{\epsilon I}{2}+\Sigma_0^{1/2}\Phi_{10}'M_{10}^{-1}\Phi_{10}\Sigma_0^{1/2}\right.\\&~~~ \left.-(\frac{\epsilon^2 I}{4}+\Sigma_0^{1/2}\Phi_{10}'M_{10}^{-1}\Sigma_1M_{10}^{-1}\Phi_{10}\Sigma_0^{1/2})^{1/2}\right]\Sigma_0^{-1/2}.
	 \end{align*}
Taking the limit as $\epsilon\rightarrow 0$ we obtain the solution to the deterministic problem, which corresponds to the initial condition 
	\begin{align*}
		\Pi_0(0)&=\Sigma_0^{-1/2}\left[\Sigma_0^{1/2}\Phi_{10}'M_{10}^{-1}\Phi_{10}\Sigma_0^{1/2}\right.\\
		&~~~ \left.-(\Sigma_0^{1/2}\Phi_{10}'M_{10}^{-1}\Sigma_1M_{10}^{-1}\Phi_{10}\Sigma_0^{1/2})^{1/2}\right]\Sigma_0^{-1/2}.
	 \end{align*}

\section{Stationary case and Invariant measure}\label{sec:stationary}
We now turn to the stationary counterpart of the Problem \ref{pro:meanfield}. We would like to design a cost function that will lead the the agents to achieve a given invariant measure $\rho$, if the agents follows the equilibrium strategy. In particular, given $N$ agents with identical dynamics \eqref{eq:dynagent} that attempt to minimize their control effort
	\[
		J_i (u_i)=\limsup_{T\rightarrow \infty} \frac{1}{T}\mE\left\{\int_0^T \frac{1}{2}\|u_i(t)\|^2 dt\right\},
	\]
we look for an extra cost $g(x,\mu)$ term added to the above such that, in the equilibrium state, the agents have some specified distribution. The new cost function is
	\begin{equation}\label{eq:coststatic}
		J_i (u_i)=\limsup_{T\rightarrow \infty} \frac{1}{T}\mE\left\{\int_0^T [\frac{1}{2}\|u_i(t)\|^2 +g(x_i(t),\mu^N(t))] dt\right\}
	\end{equation}
where $\mu^N$ is the empirical distribution \eqref{eq:empiricaldis}. Again we are interested in the mean-field limit of the problem, that is, the case when $N$ goes to $\infty$.

Let's first recall some relevant results in the stationary mean-field game problems. Suppose the $N$ agents with dynamics \eqref{eq:dynagent} attempt to minimize the cost function
	\[
		J_i (u_i)=\limsup_{T\rightarrow \infty} \frac{1}{T}\mE\left\{\int_0^T f(x_i(t),u_i(t),\mu^N(t)) dt\right\}.
	\]
We restrict ourself to equilibriums with symmetric stationary Markovian strategies 
	\[
		u_i(t)=\phi(x_i(t)).
	\] 
In the mean-field limit, one can adapt the following steps \cite{CarDelLac13}. First, fix a probability measure $\mu$ and then solve the standard stochastic control problem (parametrized by $\mu$) 
	\begin{equation}
		\phi^\star=\argmin_{\phi} \lim_{T\rightarrow \infty} \frac{1}{T}\mE\left\{\int_0^T f(x(t),\phi(x(t)),\mu(t)) dt\right\}
	\end{equation}
subject to the dynamics
	\begin{equation}\label{eq:dynamicsstatic}
		dx(t)=Ax(t)dt+\barA \xmu dt +B\phi(x(t))dt+Bdw(t).
	\end{equation}
Once this standard optimal control problem is solved, the remaining issue is finding the correct distribution $\mu$ such that the stationary distribution of \eqref{eq:dynamicsstatic} with optimal control strategy 
	\[
		u^\star(t)=\phi^\star(x(t))
	\]
coincides with $\mu$. The solution to this mean-field game problem involves the coupled PDEs \cite{LasLio07,CarDelLac13}
	   \begin{subequations}\label{eq:meanfieldgamesta}
    \begin{equation}\label{eq:meanfieldgamesta1}
       \frac{1}{2}\trace(BB'\nabla^2 \lambda)+\eta
        -H^{\rho^\star}(x,-\nabla \lambda)=0,
    \end{equation}
    \begin{equation}\label{eq:meanfieldgamesta2}
   \nabla\cdot((Ax+\barA \bar x_{\rho^\star}+B\phi^\star)\rho^\star)-\frac{1}{2}\trace(BB'\nabla^2 \rho^\star)=0,
    \end{equation}
    \begin{equation}
    \rho^\star \ge 0, ~~~\int \rho^\star =1,
    \end{equation}
    \end{subequations}
where $\eta$ is a constant, $u^\star=\phi^\star(x)$ is the minimizer of
	\[
		H^\rho(x,p)=\min_{u\in\mR^m}\left\{p^\prime (Ax+\barA \bar x_{\rho}+Bu)+f(x,u,\rho)\right\}.
	\] 
When the cost function is of the form \eqref{eq:coststatic}, the PDEs boil down to

    \begin{subequations}\label{eq:meanfieldgamestan}
    \begin{equation}\label{eq:meanfieldgamestan1}
        \frac{1}{2}\trace(BB'\nabla^2 \lambda)+\eta+\nabla\lambda\cdot Ax+\nabla\lambda\cdot\barA \bar x_{\rho^\star}
        +\frac{1}{2}\nabla\lambda\cdot BB'\nabla\lambda-g(x,\rho^\star)=0,
    \end{equation}
    \begin{equation}\label{eq:meanfieldgamestan2}
   \nabla\cdot((Ax+\barA \bar x_{\rho^\star}+BB'\nabla\lambda)\rho^\star)-\frac{1}{2}\trace(BB'\nabla^2 \rho^\star)=0,
    \end{equation}
    \begin{equation}\label{eq:meanfieldgamestan3}
    \rho^\star \ge 0, ~~~\int \rho^\star =1.
    \end{equation}
    \end{subequations}
The existence of a solution $(\rho^\star, \lambda)$ can be shown under some proper assumptions on $g$, see \cite{LasLio07,CarDelLac13}. 

Back to our problem, the cost function $g$ in \eqref{eq:coststatic} becomes a design parameter, which is different from the classic MFG setting. Our goal is to choose a function $g$ such that the corresponding stationary distribution in Nash equilibrium is $\rho^\star$. The solution relies on the same PDEs \eqref{eq:meanfieldgamestan}, but with different variables. Given a distribution $\rho^\star$, we need to find $\lambda$ and the proper cost $g$ that solve \eqref{eq:meanfieldgamestan}. It turns out that \eqref{eq:meanfieldgamestan} has solution only for a small class of distributions $\rho^\star$, which we call the feasible distributions. 
We next focus on Gaussian distributions. In this case, the feasible distributions can be characterized by some algebraic equations. The cases of general distributions will be investigated in future study.

Let $\rho^\star$ be a Gaussian distribution with mean $m$ and covariance $\Sigma$. Plugging the ansatz  
	\[
		\lambda(x)=-\frac{1}{2}x'\Pi x+n'x
	\]
with $\Pi=\Pi'$, into \eqref{eq:meanfieldgamestan} yields (after discarding constant terms)
	\begin{subequations}\label{eq:gaussian}
	\begin{eqnarray}\label{eq:gaussian1}
	&&-\frac{1}{2}\trace(BB'\Pi)+\eta+\frac{1}{2}x'(-\Pi A-A'\Pi+\Pi BB'\Pi)x\\&&+n'(A-BB'\Pi)x-m'\barA '\Pi x-g(x,\rho^\star)=0\nonumber
	\end{eqnarray}
	\begin{equation}\label{eq:gaussian2}
	(A-BB'\Pi)\Sigma+\Sigma(A-BB'\Pi)'+BB'=0
	\end{equation}
	\begin{equation}\label{eq:gaussian3}
	(A+\barA -BB'\Pi)m+BB'n=0.
	\end{equation}
	\end{subequations}
In order for the solution to exist, in view of \eqref{eq:gaussian2}, it is necessary that $\Sigma$ satisfies 
	\begin{subequations}\label{eq:algcondition}
	\begin{equation}\label{eq:algcondition1}
		A\Sigma+\Sigma A' \in \range (\f_B),
	\end{equation}
where $\f_B(X)=BX'+XB'$ is a map from $\mR^{n\times m}$ to the space of symmetric matrices (see \cite{CheGeoPav14b} for other equivalent algebraic conditions). Likewise, by \eqref{eq:gaussian3}, the mean $m$ has to satisfy
	\begin{equation}\label{eq:algcondition2}
		(A+\bar A)m \in \range(B).
	\end{equation}
	\end{subequations}
On the other hand, given $(m,\Sigma)$ satisfying \eqref{eq:algcondition}, assuming $B$ has full column rank, then \eqref{eq:gaussian2} has a unique symmetric solution \cite{CheGeoPav14b}. Therefore, these two conditions are also sufficient. Now from \eqref{eq:gaussian1} it is easy to conclude that a possible cost function is
	\begin{subequations}\label{eq:stadesigncost}
	\begin{equation}
		g(x,\rho)=\frac{1}{2}x'Qx+n\cdot(A-BB'\Pi)x-\bar x_\rho\cdot \barA '\Pi x,
	\end{equation}
with 
	\begin{equation}
		Q=-\Pi A-A'\Pi+\Pi BB'\Pi,
	\end{equation}
	\end{subequations}
with $\Pi$ being the unique solution to \eqref{eq:gaussian2}.
Therefore, we have established the following result.
\begin{thm}
Consider $N$ agents governed by \eqref{eq:dynagent}. Suppose the $g$ function in the cost \eqref{eq:coststatic} is as in \eqref{eq:stadesigncost}, then, in the Nash equilibrium, the agents will reach stationary Gaussian distribution with mean $m$ and covariance $\Sigma$, in the limit as $N$ goes to $\infty$.
\end{thm}

\section{Cooperative game}\label{sec:coop}
In this section we shift to a slightly different problem. Given the same interacting agents' system \eqref{eq:dynagent}, we would like to investigate the density steering problem in the cooperative game setting. How to select an optimal controller to drive the agents from given initial distribution $\rho_0$ to terminal distribution $\rho_1$? Again, we restrict ourself to equilibriums given by symmetric Markovian strategies in closed-loop feedback form
	\begin{equation}\label{eq:symstrategy1}
		u_i(t)=\phi(t,x_i(t)), i=1,\ldots,N.
	\end{equation}
The cost function we attempt to minimize is the average control energy
	\begin{equation}\label{eq:controlenergy}
	J(u)=\mE\left\{\sum_{i=1}^N\int_0^1 \frac{1}{2}\|u_i(t)\|^2 dt\right\}.
	\end{equation}
We are interested in the mean-field limit, namely, the asymptotical behavior of the solution when $N\rightarrow \infty$.
\begin{problem}\label{pro:interacting}
Given $N$ agents governed by \eqref{eq:dynagent} with initial density $\rho_0$, find a control strategy \eqref{eq:symstrategy1} with minimum control energy \eqref{eq:controlenergy} so that the agents will reach density $\rho_1$ at time $t=1$, as $N$ goes to $\infty$.
\end{problem}

The major difference between this problem and the mean-field game is that all the agents always use the same control strategy. A small perturbation on the control will affect the probability density flow as the perturbation is applied to the controllers of all the agents, see \cite{CarDelLac13,CarDel13} for more discussions on their differences. The average control energy \eqref{eq:controlenergy} is equivalent to relative entropy of the controller system with respect to the uncontrolled system \cite{DawGar87,Fen06,Fis14}. Therefore, the above problem can also be viewed as an Schr\"odinger bridge problem for interacting particle systems.

Problem \ref{pro:interacting} can be formulated as an optimal control problem over the McKean-Vlasov model
    \begin{equation}\label{eq:controlleddynamics}
		dx(t)=Ax(t)dt+\barA \bar{x}(t) dt +Bu(t)dt+Bdw(t), ~~x(0)=x_0\sim \rho_0.
	\end{equation}
It has the following fluid dynamic formulation. Let $\rho(t,\cdot)$ be the probability density of the controlled process $x(t)$, then the optimal control problem can be stated as
    \begin{subequations}\label{eq:optimization}
    \begin{align}
    & \inf_{\rho, u}\quad \int_0^1 \int_{\mR^n} \frac{1}{2} \rho(t,x) \|u(t,x)\|^2 dxdt, \\
    & \frac{\partial \rho}{\partial t}+\nabla\cdot((Ax+\barA \bar x_\rho+Bu)\rho)-\frac{1}{2}\trace(BB'\nabla^2 \rho)=0, \\
    & \rho(0,\cdot)=\rho_0, \quad \rho(1,\cdot)=\rho_1.
    \end{align}
    \end{subequations}

\begin{prop}\label{prop:necessary}
 If there exists $(\lambda,\rho^\star)$ satisfying
  \begin{subequations}\label{eq:necessaryi}
    \begin{equation}\label{eq:necessaryi1}
        \frac{\partial \lambda}{\partial t}+\nabla\lambda\cdot Ax+\nabla\lambda\cdot\barA \bar x_{\rho^\star}+\frac{1}{2}\trace(BB'\nabla^2 \lambda)+\frac{1}{2}\nabla\lambda\cdot BB'\nabla\lambda+\barA x\cdot\langle \nabla\lambda, \rho^\star\rangle=0,
    \end{equation}
    \begin{equation}\label{eq:necessaryi2}
    \frac{\partial \rho^\star}{\partial t}+\nabla\cdot((Ax+\barA \bar x_{\rho^\star}+BB'\nabla\lambda)\rho^\star)-\frac{1}{2}\trace(BB'\nabla^2 \rho^\star)=0,
    \end{equation}
    with boundary conditions
    \begin{equation}\label{eq:necessaryi3}
    \rho^\star(0,\cdot)=\rho_0, \quad \rho^\star(1,\cdot)=\rho_1,
    \end{equation}
    \end{subequations}
then $(\rho^\star, u^\star=B'\nabla\lambda)$ is a solution to \eqref{eq:optimization}.
\end{prop}

Equations \eqref{eq:necessaryi} are highly coupled. In general, one may not expect a solution to exist. But interestingly, as we will see below, \eqref{eq:necessaryi} always has a solution. In fact, we are going to construct a solution based on standard Schr\"odinger bridge theory.

Let $(\hat\rho,\hat\lambda)$ be as in \eqref{eq:dynamics1}-\eqref{eq:zeromeancontrol},
    \begin{subequations}\label{eq:ym}
    \begin{equation}
        m(t)=\bar\Phi(1,t)'\hat M_{10}^{-1}(\bar x_{\rho_1}-\bar\Phi_{10}\bar x_{\rho_0})
    \end{equation}
and $y(t)$ the solution to
    \begin{equation}
        \dot{y}(t)=(A+\barA )y(t)+BB'm(t), ~~y(0)=\bar x_{\rho_0},
    \end{equation}
    \end{subequations}
where $\hat M_{10}=\hat M(1,0)$ with
    \[
        \hat M(t,s)=\int_s^t \bar\Phi(t,\tau)BB'\bar\Phi(t,\tau)'d\tau.
    \]
Define
	\[
		\gamma(t)=-\int_0^t (\barA y(s)\cdot m(s)+\frac{1}{2}m(s)\cdot BB'm(s))ds,
	\]
    \begin{subequations}\label{eq:optimalpairi}
    \begin{equation}\label{eq:optimalpairi1}
        \lambda(t,x)=\hat\lambda(t,x-y(t))+m(t)\cdot x+\gamma(t),
    \end{equation}
    and
    \begin{equation}\label{eq:optimalpairi2}
        \rho^\star(t,x)=\hat\rho(t,x-y(t)),
    \end{equation}
    \end{subequations}
then $(\lambda, \rho^\star)$ solves \eqref{eq:necessaryi}. Apparently, \eqref{eq:optimalpairi2} satisfies the boundary conditions \eqref{eq:necessaryi3}. To verify \eqref{eq:necessaryi2}, substitute \eqref{eq:optimalpairi} into \eqref{eq:necessaryi2}, which gives
    \begin{eqnarray*}
        &&\frac{\partial \rho^\star}{\partial t}+\nabla\cdot((Ax+\barA \bar x_{\rho^\star}+BB'\nabla\lambda)\rho^\star)-\frac{1}{2}\trace(BB'\nabla^2 \rho^\star)
        \\&=& \frac{\partial \hat\rho}{\partial t}-\nabla\hat\rho\cdot((A+\barA )y+BB'm)+\nabla\cdot((Ax+BB'\nabla\lambda)\hat\rho)\\&&~~~ +\barA \langle \xi, \hat\rho(\xi-y)\rangle\cdot \nabla \hat\rho-\frac{1}{2}\trace(BB'\nabla^2 \hat\rho)
        \\&=&-\nabla\hat\rho\cdot(\barA y+BB'm)+\nabla\cdot (BB'm\hat\rho)+\barA \langle\xi, \hat\rho(\xi-y)\rangle\cdot\nabla\hat\rho=0.
    \end{eqnarray*}
 Similarly, Combing \eqref{eq:optimalpairi} and \eqref{eq:necessaryi1} yields
    \begin{eqnarray*}
    &&\frac{\partial \lambda}{\partial t}+\nabla\lambda\cdot Ax+\nabla\lambda\cdot\barA \bar x_{\rho^\star}+\frac{1}{2}\trace(BB'\nabla^2 \lambda)+\frac{1}{2}\nabla\lambda\cdot BB'\nabla\lambda+\barA x\cdot\langle \nabla\lambda, \rho^\star\rangle
    \\&=& \frac{\partial \hat\lambda}{\partial t}-\nabla\hat\lambda\cdot((A+\barA )y+BB'm)\\&&~~~ +\dot{m}\cdot x+\dot{\gamma}+(\nabla\hat\lambda+m)\cdot (Ax+\barA \bar x_{\rho^\star})+\frac{1}{2}\trace(BB'\nabla^2 \hat\lambda)
    \\&&+\frac{1}{2}(\nabla\hat\lambda+m)\cdot BB'(\nabla\hat\lambda+m)+\barA x\cdot\langle \nabla\hat\lambda+m, \rho^\star\rangle
    \\&=& -\nabla\hat\lambda\cdot\barA y+\dot{m}\cdot x+\dot{\gamma}+(\nabla\hat\lambda+m)\cdot\barA \bar x_{\rho^\star}\\&&~~~+m\cdot Ax+\frac{1}{2}m\cdot BB'm+\barA x\cdot m+\barA x\cdot\langle \nabla\hat\lambda, \rho^\star\rangle
    \\&=& \barA x\cdot\langle \nabla\hat\lambda(\xi-y), \hat\rho(\xi-y)\rangle=0.
    \end{eqnarray*}
Therefore, the pair $( \rho^\star, u^\star=B'\nabla\lambda)$ is indeed a solution to \eqref{eq:necessaryi}. Next we prove that this pair $(\rho^\star, u^\star)$ provides a solution to the optimal control problem \eqref{eq:optimization}.

Let $\bar{u}(t)=\mE\{u(t)\}$, then, by \eqref{eq:controlleddynamics}, we have
    \begin{equation}\label{eq:meancontrol}
        d\bar{x}(t)=(A+\barA )\bar{x}(t)dt+B\bar{u}(t)dt
    \end{equation}
and
    \begin{equation}\label{eq:variance}
        d(\tilde{x}(t))=A\tilde{x}(t)dt+B\tilde{u}(t)dt+Bdw,
    \end{equation}
where $\tilde x=x-\bar x$ and $\tilde u=u-\bar u$.
The control energy can then be decomposed into two parts as
    \begin{equation*}
        \mE\{\int_0^1 \frac{1}{2} \|u(t)\|^2dt\} =  \int_0^1 \frac{1}{2}\|\bar{u}(t)\|^2 dt+\mE\{\int_0^1 \frac{1}{2} \|\tilde u(t)\|^2dt\}.
    \end{equation*}
These two parts of control energy, corresponding to $\bar{u}$ and  $u-\bar{u}$ respectively, can be minimized independently since the dynamics \eqref{eq:meancontrol} and \eqref{eq:variance} are decoupled. We next show that\\
i) $\bar{u}^\star$ minimizes
    \begin{equation}\label{eq:meanenergy}
        \int_0^1 \frac{1}{2}\|\bar{u}(t)\|^2 dt
    \end{equation}
ii) $\tilde u^\star$ minimizes
    \begin{equation}\label{eq:varianceenergy}
        \mE\{\int_0^1 \frac{1}{2} \|\tilde u(t)\|^2dt\}.
    \end{equation}
For i), recalling
    \[
        u^\star(t)=B'\nabla \lambda=B'\nabla\hat\lambda(t,x(t)-y(t))+B'm(t),
    \]
we have
    \begin{align*}
        \bar{u}^\star(t)&=B'\mE\{\nabla\hat\lambda(t,x(t)-y(t))\}+B'm(t)\\
&        =B'\langle \nabla \hat\lambda(t,x-y(t)), \rho^\star(t,x)\rangle+B'm(t)=B'm(t).
    \end{align*}
Using standard optimal control, it is easy to see that $\bar{u}^\star(t)$ minimizes \eqref{eq:meanenergy} subject to \eqref{eq:meancontrol} and boundary conditions
    \[
        \bar{x}(0)=\bar x_{\rho_0},~~ \bar{x}(1)=\bar x_{\rho_1}.
    \]
We next show ii). Note
    \[
        \tilde u^\star(t)=B'\nabla\hat\lambda(t,x(t)-y(t))=B'\nabla\hat\lambda(t,x(t)-\bar{x}(t))=
        B'\nabla\hat\lambda(t,\tilde x(t)).
    \]
By Schr\"odinger bridge theory \cite{CheGeoPav14e,CheGeoPav15b} (see also \eqref{eq:SB1}-\eqref{eq:lambdahat}), it minimizes
    $$\mE\{\int_0^1 \frac{1}{2}\|\tilde{u}(t)\|^2dt\}$$
subject to
    \[
        d\tilde{x}(t)=A\tilde{x}(t)dt+B\tilde{u}(t)dt+Bdw
    \]
and marginal distributions of $\tilde{x}(0)$ and $\tilde{x}(1)$, which are $\hat\rho_0$ and $\hat\rho_1$, respectively. Hence we have established the following result.
\begin{thm}\label{thm:SBi}
The pair $(\rho^\star, u^\star=B'\nabla\lambda)$ with $\rho^\star, \lambda$ as in \eqref{eq:optimalpairi} solves \eqref{eq:optimization}.
\end{thm}

\subsection{Linear-Quadratic case}
When both of the marginals $\rho_0$ and $\rho_1$ are normal distributions, the optimal control $u^\star$ is a linear state-feedback control.
Let the two marginal distributions $\rho_0$ and $\rho_1$ be
    \[
        \rho_0 \sim \cN[m_0, \Sigma_0],~~\rho_1\sim \cN[m_1, \Sigma_1].
    \]
By Theorem \ref{thm:SBi}, we need only to compute $\lambda$ as in \eqref{eq:optimalpairi1}, which is
    \[
        \lambda(t,x)=\hat\lambda(t,x-y(t))+m(t)\cdot x+\gamma(t).
    \]
The function $\hat\lambda$ corresponds to the Schr\"odinger bridge \eqref{eq:SB1}, which satisfies \cite{CheGeoPav14a}
    \[
        \nabla\hat\lambda(t,x)=-\Pi(t)x,
    \]
where $\Pi(t)$ is the solution to the Riccati equation
    \[
        \dot{\Pi}(t)=-A'\Pi(t)-\Pi(t)A+\Pi(t)BB'\Pi(t)
    \]
with boundary condition
    \begin{align*}
        \Pi(0)&=\Sigma_0^{-1/2}[\frac{I}{2}+\Sigma_0^{1/2}\Phi_{10}'M_{10}^{-1}\Phi_{10}\Sigma_0^{1/2}\\&~~~ -(\frac{I}{4}+\Sigma_0^{1/2}\Phi_{10}'M_{10}^{-1}\Sigma_1M_{10}^{-1}\Phi_{10}\Sigma_0^{1/2})^{1/2}]\Sigma_0^{-1/2}.
    \end{align*}
It follows the optimal control is
    \[
        u^\star(t)=B'\nabla\lambda(t,x(t))=-B'\Pi(t)(x(t)-y(t))+B'm(t)=-B'\Pi(t)x(t)+B' n(t),
    \]
where
    \begin{eqnarray*}
        n(t)&=&\Pi(t)y(t)+m(t)
        \\&=&\Pi(t)\bar\Phi(t,1)\hat M(1,t)\hat{M}_{10}^{-1}\bar\Phi_{10}m_0\\
        &&~~~ +\Pi(t)\hat{M}(t,0)\bar\Phi(1,t)'\hat{M}_{10}^{-1}m_1
        +\bar\Phi(1,t)'\hat M_{10}^{-1}(m_1-\bar\Phi_{10}m_0).
    \end{eqnarray*}

\subsection{Zero-noise limit}
We study the optimal steering problem for McKean-Vlasov model \eqref{eq:controlleddynamics} with smaller disturbance $\sqrt{\epsilon} dw(t)$, namely,
	\begin{equation}\label{eq:dynamicsreduce}
		dx(t)=Ax(t)dt+\barA \bar{x}(t) dt +Bu(t)dt+\sqrt{\epsilon}Bdw(t), ~~x(0)=x_0~ \mbox{a.s.}.
	\end{equation}
In particular, we are interested in the zero-noise limit of this problem. That is, optimal steering problem for dynamics
	\begin{equation}\label{eq:dynamicszero}
		dx(t)=Ax(t)dt+\barA \bar{x}(t) dt +Bu(t)dt, ~~x(0)=x_0~ \mbox{a.s.}.
	\end{equation}
We show that the probability flow of the solution to the latter is the limit of that of the former as $\epsilon$ goes to $0$. This is achieved in a constructive manner.

Let's start with the steering problem for the dynamics
	\[
		dx(t)=Ax(t)dt+Bu(t)dt
	\]
with marginals $\hat{\rho}_0$ and $\hat{\rho}_1$. This problem has been studied in \cite{CheGeoPav15b} and the solution is 
	\begin{equation}\label{eq:optimalcontrolzero}
		\tilde u(t,x)=B'\Phi(1,t)'M_{10}^{-1}(T\circ T_t^{-1}(x)-\Phi_{10}(T_t^{-1}(x)))
	\end{equation}
where $T$ is the generalized optimal mass transport map \cite{CheGeoPav15b} with marginals $\hat{\rho}_0, \hat{\rho}_1$, and 
	\[
		T_t(x)=\Phi(t,1)M(1,t)M_{10}^{-1}\Phi_{10}x+M(t,0)\Phi(1,t)'M_{10}^{-1}T(x).
	\]
The corresponding distribution flow is 
	\[
	\hat\rho(t,\cdot)=(T_t)_\sharp \hat\rho_0.
	\]
Note that $\hat\rho$ and $\tilde{u}$ satisfy 
the continuity equation
	\[
		\frac{\partial \hat\rho}{\partial t}+\nabla\cdot((Ax+B\tilde{u})\hat\rho)=0.
	\]
We claim that 
	\begin{equation}\label{eq:optimalcontrolzero1}
		u^\star(t,x)=\tilde{u}(t,x-y(t))+B'm(t)
	\end{equation}
with $y, m$ in \eqref{eq:ym}, is the optimal control strategy for the steering problem for the dynamics \eqref{eq:dynamicszero} and the corresponding distribution flow is
	\begin{equation}\label{eq:flowOMT}
		\rho^\star(t,x)=\hat{\rho}(t,x-y(t)).
	\end{equation}
We shall skip the proof as it is similar to the case with disturbance (see \eqref{eq:optimalpairi}-\eqref{eq:varianceenergy}).

The solution to the density steering problem with dynamics \eqref{eq:dynamicsreduce}, weakly converges to $\rho^\star$ as $\epsilon$ goes to $0$. This follows directly from the fact a Schr\"odinger bridge converges to the corresponding  optimal transport solution \cite{CheGeoPav15b} as $\epsilon$ goes to $0$.

 
\section{Examples}\label{sec:example}
Consider $N$ agents with dynamics
    \[
        dx_i(t)=x_i(t)dt-\frac{2}{N-1}\sum_{j\neq i}x_j(t)dt+dw_i(t), ~1\le i\le N.
    \]
The two marginal distributions $\rho_0$ and $\rho_1$ are two normal distributions
    \[
        \rho_0 \sim \cN[1, 4],\quad \rho_1 \sim \cN[-4, 1].
    \]
\subsection{Noncooperative game}
\begin{figure}
  \centering
  \includegraphics[width=0.6\textwidth]{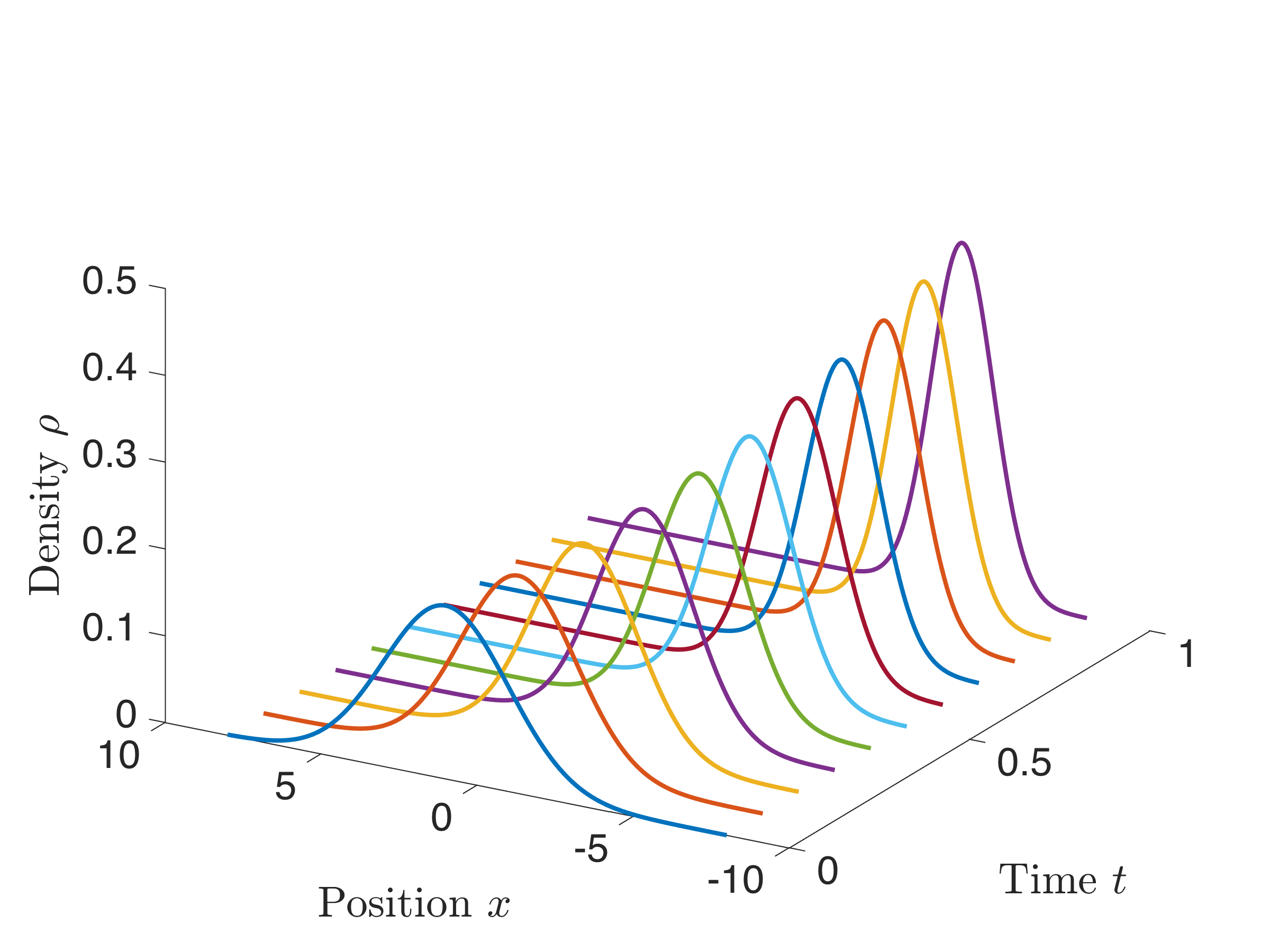}\\
  \caption{Time evolution of probability densities}\label{fig:Mrho}
\end{figure}

\begin{figure}
  \centering
  \includegraphics[width=0.6\textwidth]{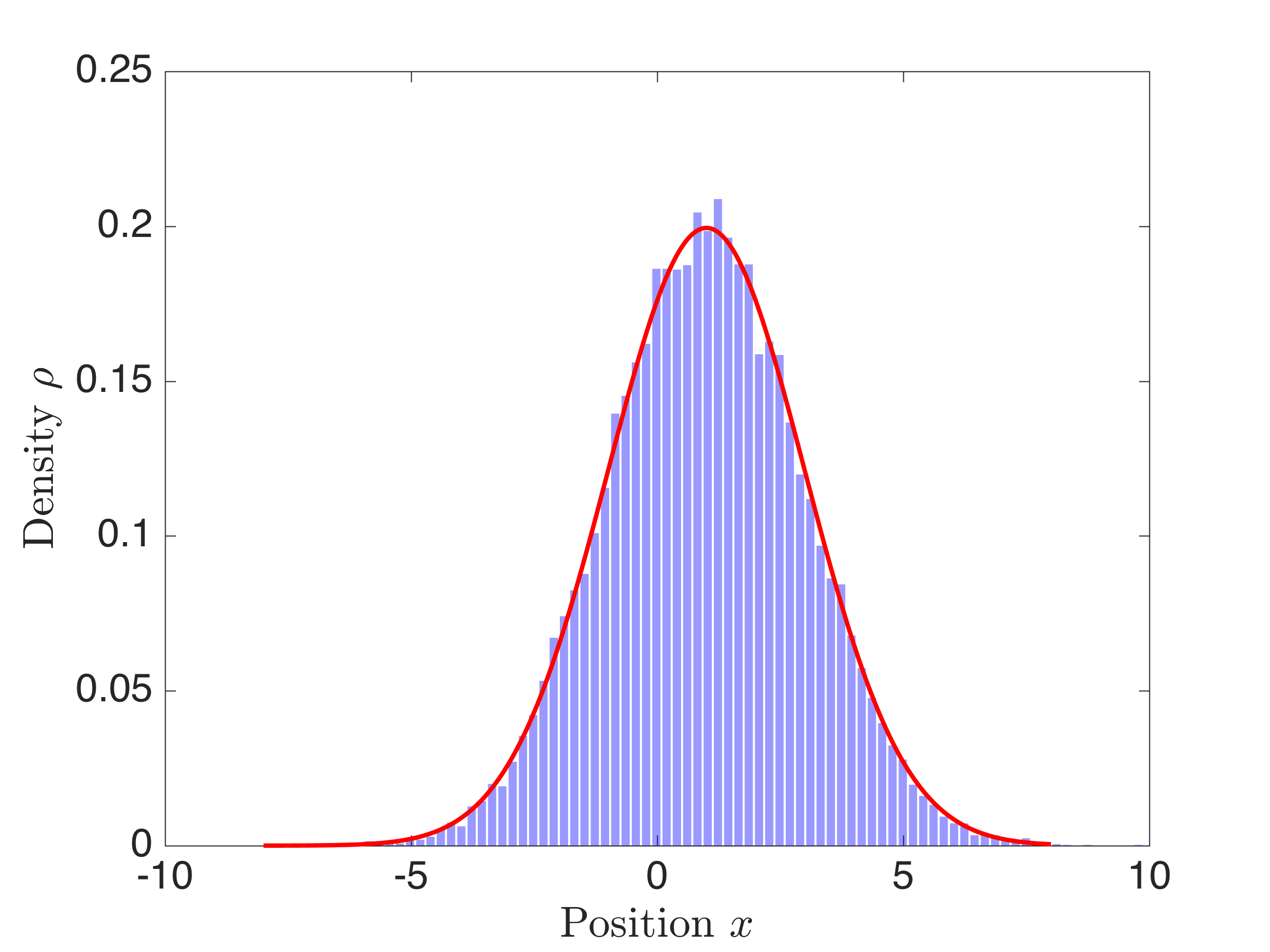}\\
  \caption{Empirical distribution of $x(0)$}\label{fig:Mrho0}
\end{figure}

\begin{figure}
  \centering
  \includegraphics[width=0.6\textwidth]{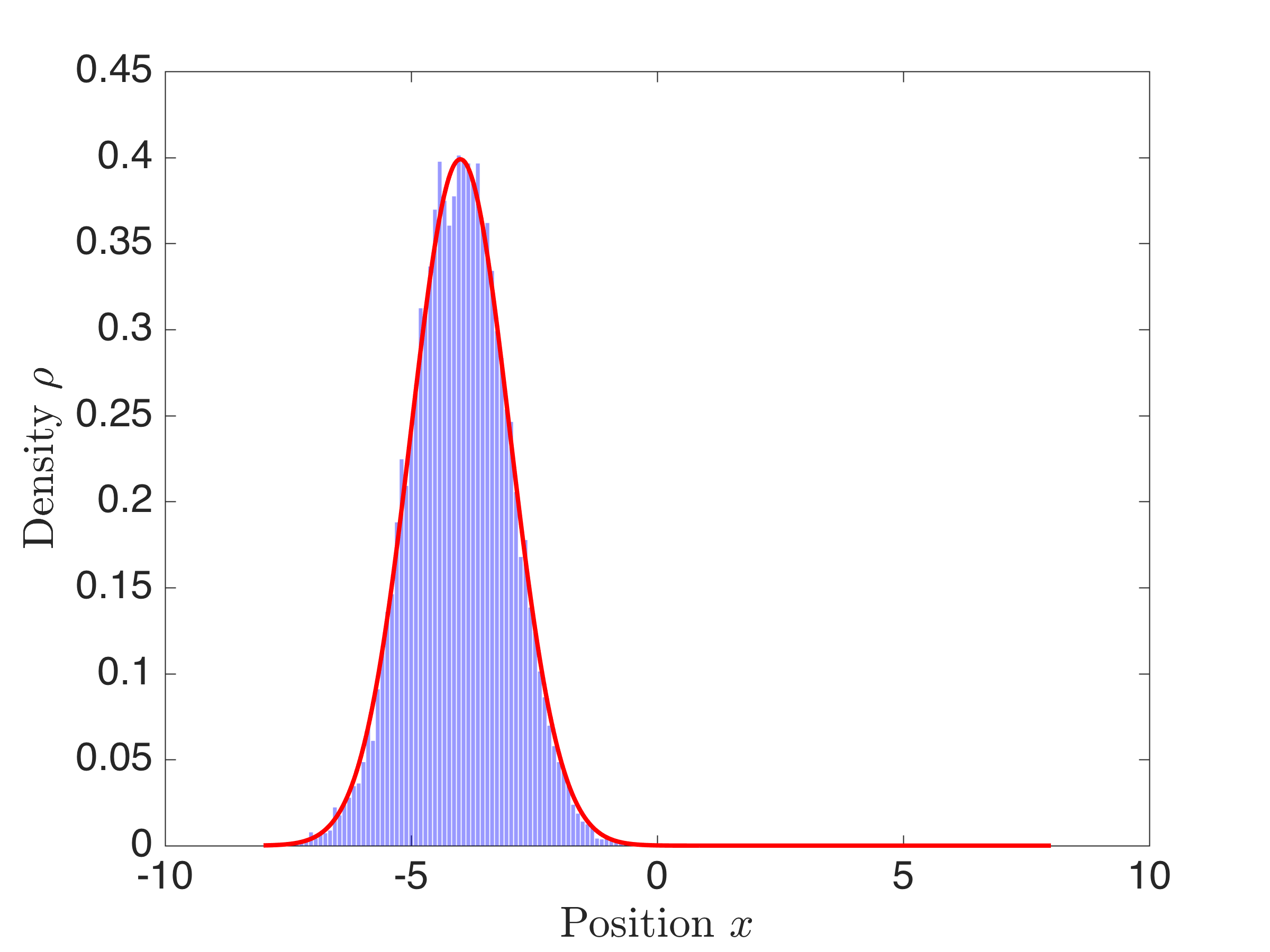}\\
  \caption{Empirical distribution of $x(1)$}\label{fig:Mrho1}
\end{figure}

\begin{figure}
  \centering
  \includegraphics[width=0.6\textwidth]{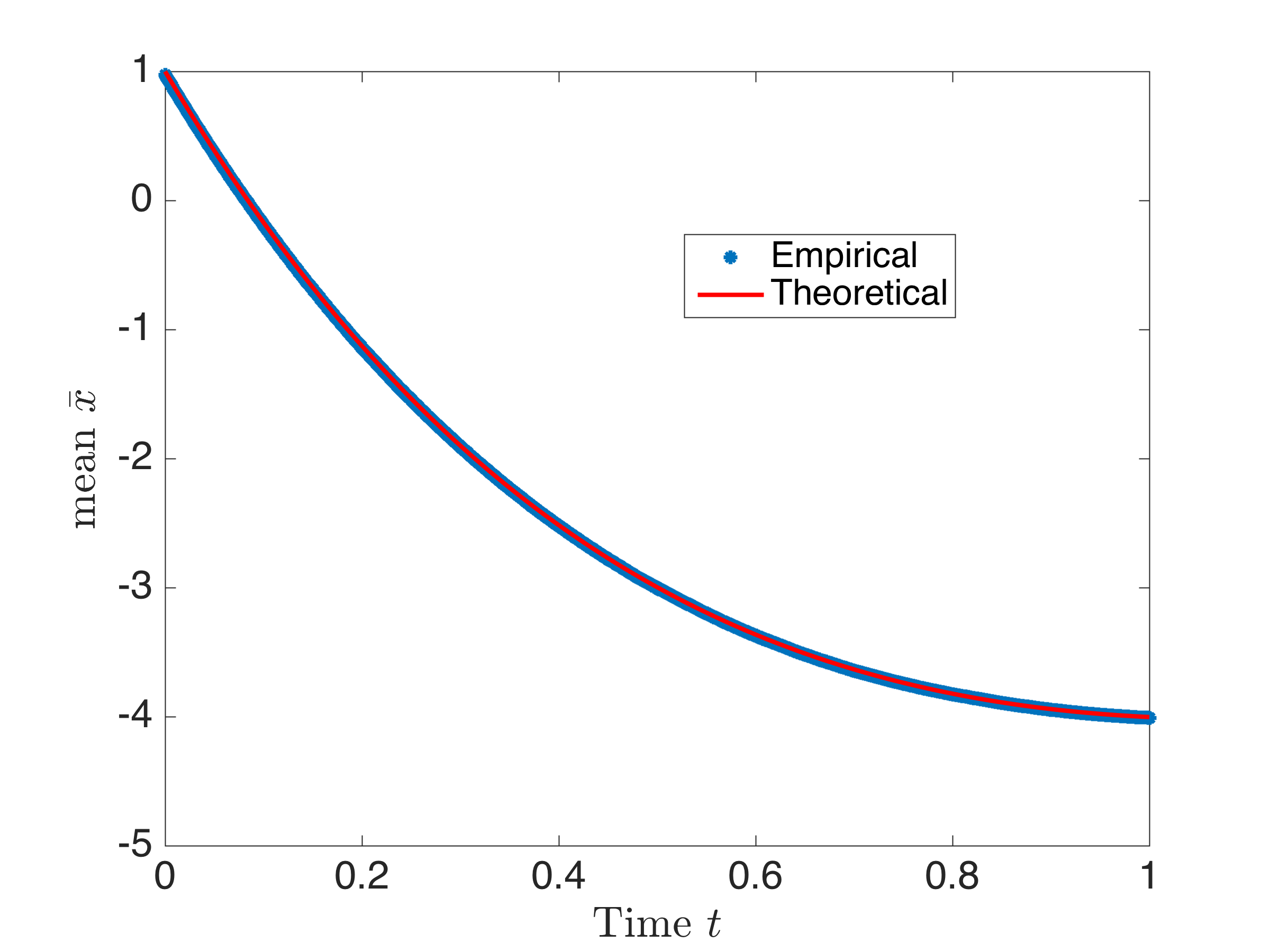}\\
  \caption{Time evolution of mean $\bar{x}(t)$}\label{fig:Mmean}
\end{figure}

One choice of terminal cost that will steer the agents from $\rho_0$ to $\rho_1$ is
	\[
		g(x,\mu)=0.9805(x-\xmu)^2+4.3679x.
	\]
Figure \ref{fig:Mrho} showcases the evolution of the probability density in the Nash equilibrium.
To show that the distribution of the agents would evolve according to Figure \ref{fig:Mrho}, we simulated the dynamics for a system with $N=20000$ agents under the optimal strategy. Figure \ref{fig:Mrho0} and Figure \ref{fig:Mrho1} depict the empirical distributions of the particles at time $t=0$ and $t=1$. They match with the theoretical distributions $\rho_0$ and $\rho_1$ very well. We also show the empirical mean of these particles in Figure \ref{fig:Mmean}, which perfectly matches the theoretical result.


\subsection{Cooperative game}
\begin{figure}
  \centering
  \includegraphics[width=0.6\textwidth]{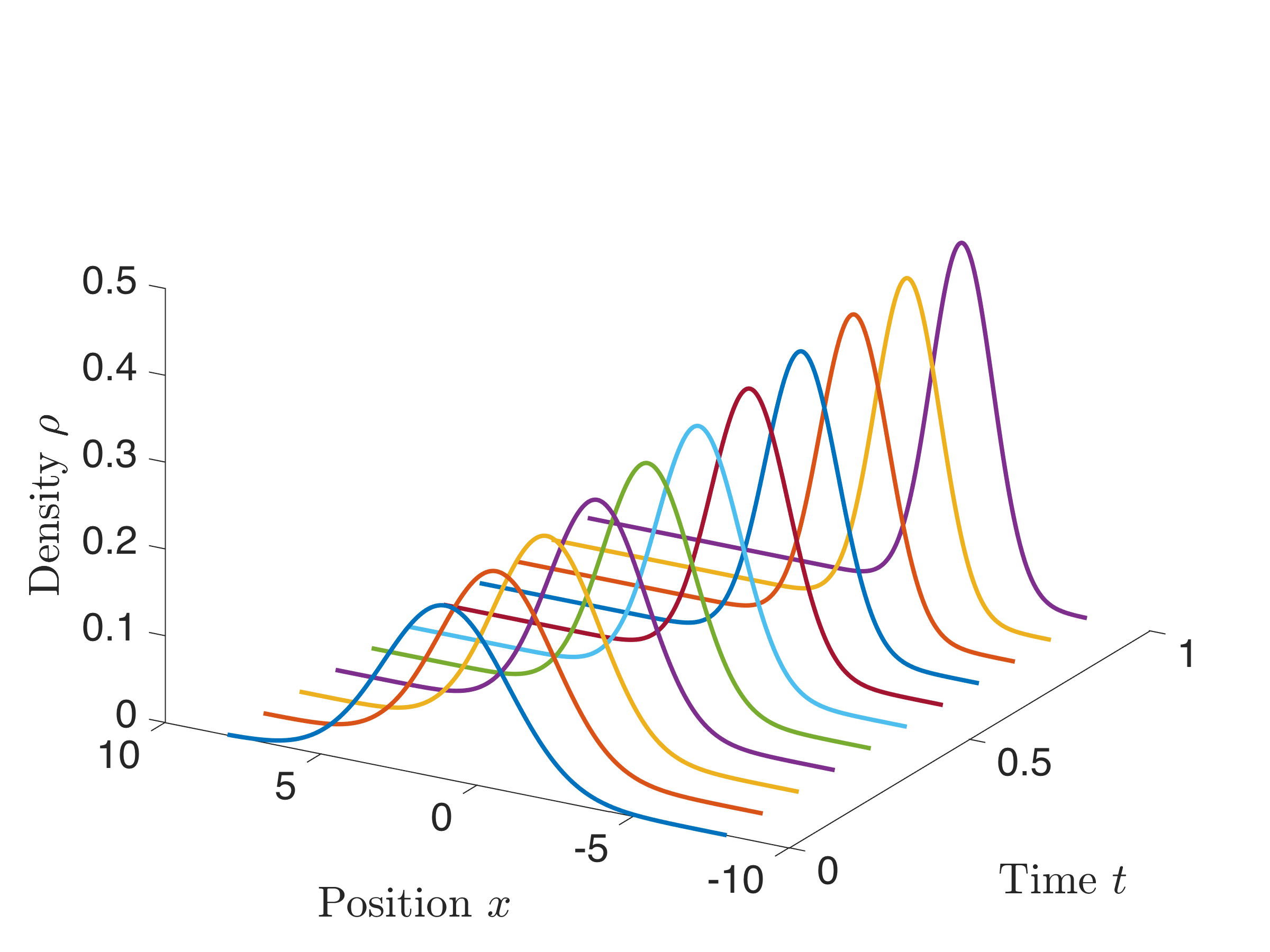}\\
  \caption{Time evolution of probability densities}\label{fig:rho}
\end{figure}

\begin{figure}
  \centering
  \includegraphics[width=0.6\textwidth]{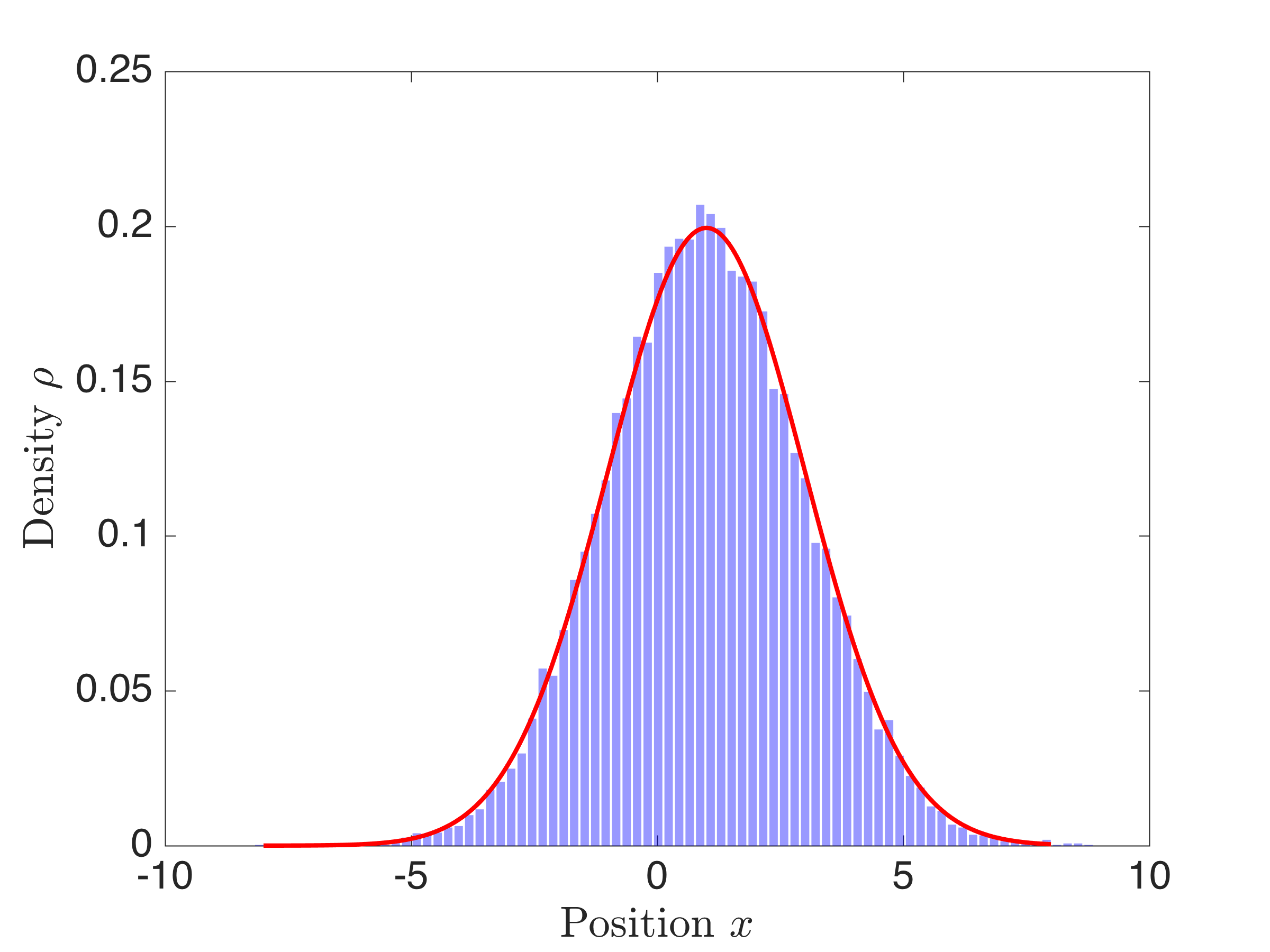}\\
  \caption{Empirical distribution of $x(0)$}\label{fig:rho0}
\end{figure}

\begin{figure}
  \centering
  \includegraphics[width=0.6\textwidth]{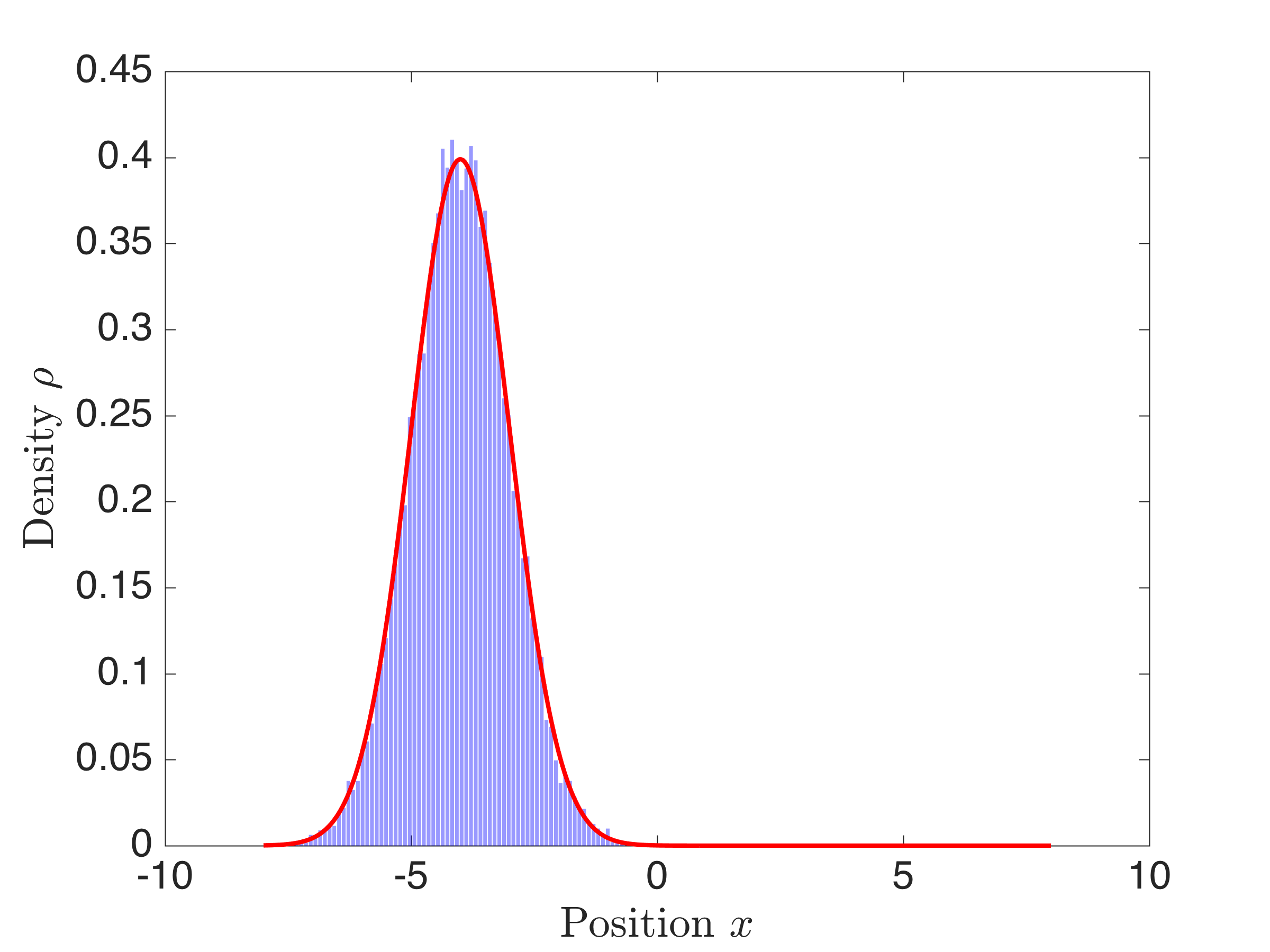}\\
  \caption{Empirical distribution of $x(1)$}\label{fig:rho1}
\end{figure}

\begin{figure}
  \centering
  \includegraphics[width=0.6\textwidth]{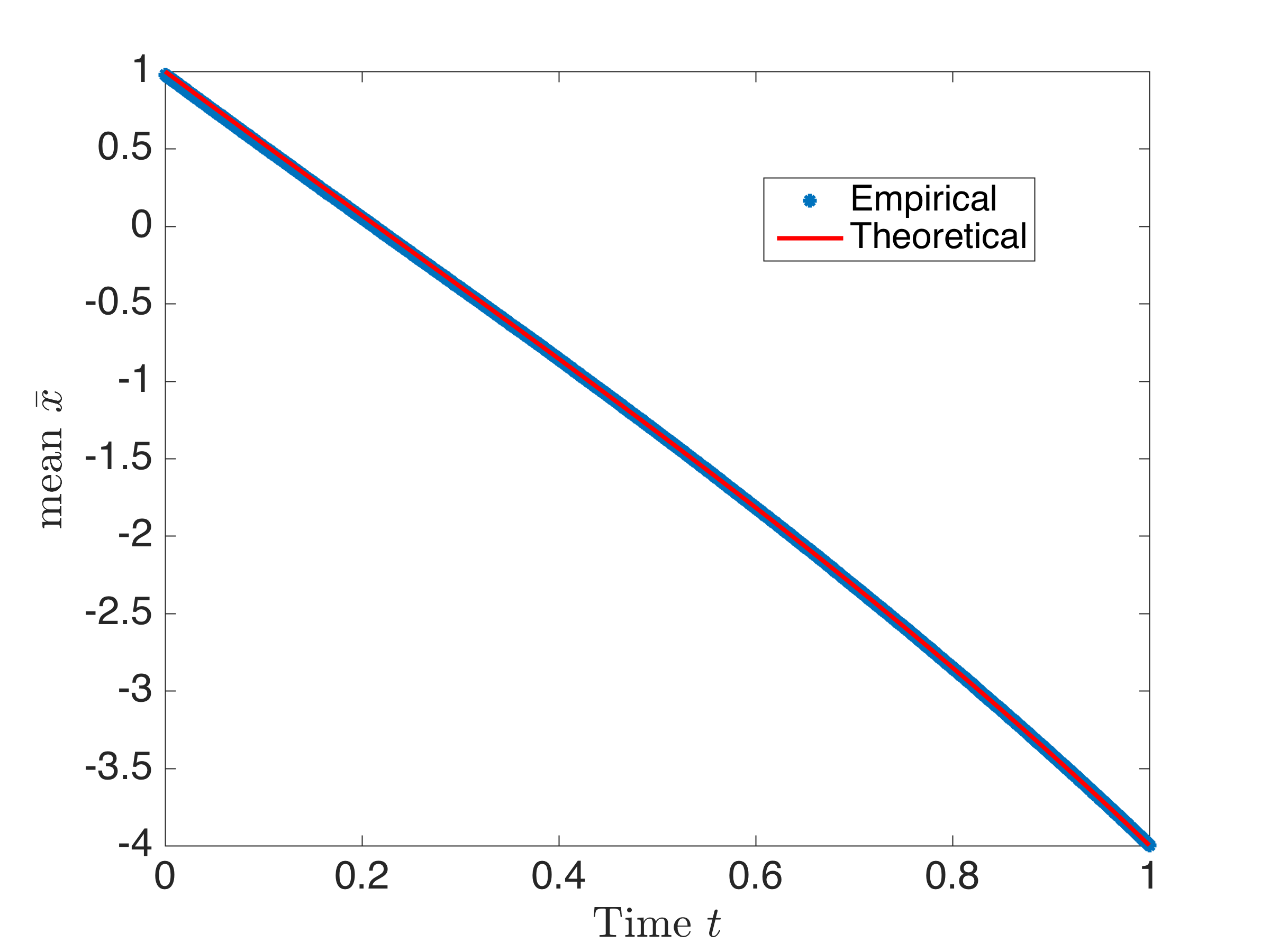}\\
  \caption{Time evolution of mean $\bar{x}(t)$}\label{fig:mean}
\end{figure}

Figure \ref{fig:rho} depicts the time evolution of the probability densities with these two marginal distributions in the cooperative game setting.
Similarly, we ran some simulations for a particle system with $N=20000$ and obtained Figure \ref{fig:rho0} and Figure \ref{fig:rho1} as the empirical distributions of the agents at time $t=0$ and $t=1$. We also show the empirical mean of these particles in Figure \ref{fig:mean}. Clearly the mean is different to the Nash equilibrium in the noncooperative game setting.

\section{Conclusion}\label{sec:conclusion}
We introduce a paradigm to steer a large number of agents from one distribution to another. The problem lies in the intersection of MFG, OMT and optimal control. We study such problems for linearly weakly interacting agents and solve the problem using tools from all these three areas. Results for several extensions such as stationary and cooperative game problems are also presented. We expect this paradigm to bring in a new dimension to the study and applications of MFG and OMT.

{
\bibliographystyle{IEEEtran}
\bibliography{./refs}
}
\end{document}